\documentclass[aps,prd,twocolumn,nofootinbib,showpacs,showkeywords,superscriptaddress]{revtex4-1}

\usepackage{amssymb}
\usepackage{graphicx}
\usepackage{amsmath, amsthm}
\usepackage{epstopdf}
\usepackage{hyperref}

\newcommand{\be}{\begin{equation}}
\newcommand{\ee}{\end{equation}}

\begin{document}

\title{Paczynski-Wiita-like potential for any static spherical 
black hole in metric theories of gravity}


\author{Valerio Faraoni}
\email{vfaraoni@ubishops.ca}
\affiliation{Physics Department and STAR Research Cluster,  
Bishop's University, 2600 College St., Sherbrooke, 
Qu\'ebec, Canada J1M~1Z7
}

\author{Shawn D. Belknap-Keet}
\email{sbelknapkeet02@ubishops.ca}
\affiliation{Physics Department, Bishop's University, 2600 
College St., Sherbrooke, Qu\'ebec, Canada J1M~1Z7
}
\author{Marianne Lapierre-L\'eonard}
\email{mlapierre12@ubishops.ca}
\affiliation{Physics Department, Bishop's University, 
2600 College St., Sherbrooke, Qu\'ebec, Canada J1M~1Z7
}
\begin{abstract} 

The pseudo-Newtonian potential of Paczynski and Wiita for 
particles orbiting a Schwarzschild black hole is 
generalized to arbitrary static and spherically symmetric 
spacetimes, including black hole solutions of alternative 
theories of gravity. In addition to being more general, our 
prescription differs substantially from a previous one in 
the literature, showing that the association of a 
pseudo-Newtonian potential even with a simple black hole 
metric is not unique.

\end{abstract}

\pacs{04.70.Bw, 04.70.-s, 04.50.Kd}

\keywords{Pseudo-Newtonian potential, alternative theories 
of gravity, black hole, accretion disk}

\maketitle

\section{Introduction}
\label{sec:1}

Black holes are central to several areas of astrophysics 
and supermassive black holes in the cores of galaxies are 
important factors in galactic evolution. Black holes in 
astrophysical environments are surrounded by matter and 
fluids which may form accretion disks orbiting around 
them. Although black holes were discovered theoretically 
just after the introduction of General Relativity, most 
(if not all) relativistic theories of gravity are believed 
to have black hole solutions. There is currently much 
interest in theories of gravity alternative to Einstein  
theory for several reasons 
\cite{Willbook, Willupdate, cosmotestGR, reviews, Salvbook, 
Padilla, 
Bertietal2013}.  From the 
theoretical point of 
view, attempts to renormalize General Relativity and to 
produce a quantum theory of gravity invariably produce 
deviations from Einstein theory in the form of higher 
order derivatives in the field equations, extra degrees of 
freedom, or scalar fields coupled nonminimally to gravity 
and matter \cite{Padilla}. The present acceleration of the 
universe 
discovered with type~Ia supernovae \cite{SN} can be 
explained in the context of General Relativity by 
re-introducing the problematic cosmological constant with 
an incredible amount of fine-tuning, by advocating a 
completely {\em ad hoc} dark energy 
\cite{AmendolaTsujikawabook}, an even more 
problematic backreaction of cosmological inhomogeneities on 
the cosmic dynamics \cite{backreaction}, or by changing the 
theory of gravity altogether \cite{reviews, Salvbook, 
AmendolaTsujikawabook}. Recently, this last possibility has 
motivated an enormous interest in alternative theories of 
gravity.

Going back from cosmology to astrophysics, modifying 
gravity would have implications for black holes and 
particles and fluids surrounding them and forming accretion 
disks. There is, therefore, interest in using 
observations of black holes in the near future to test 
deviations from General Relativity and possibly 
detect scalar hair \cite{EHT, Padilla, 
Bertietal2013, scalarhair}.

Due to the complication of relativistic motions around 
black holes, pseudo-Newtonian potentials have been used for 
decades to provide an effective simplified description of 
timelike geodesics for massive particles orbiting black 
holes \cite{PW, Kovar, Marek, pseudohistorical, 
TejedaRosswog14}. These pseudo-potentials are surprisingly 
accurate in determining orbits, given their simplicity: for example, the phase 
space of massive test particles in the Schwarzschild metric 
is not too dissimilar from that of the associated  
pseudo-potential \cite{phasespace}, a fact that can be 
understood by noting that the pseudo-potential is defined 
in such a way as to preserve the fixed points of the 
relevant dynamical system \cite{PW, Marek}.

 The first pseudo-Newtonian potential 
introduced in the literature, the Paczynski-Wiita potential 
\cite{PW, Marek}, reproduces exactly the location of the 
innermost stable circular orbit (ISCO or marginally stable 
orbit) $R_\text{{\small ISCO}}$, the location of the 
marginally bound orbit $R_\text{mb}$, and the form of the 
Keplerian angular momentum $L(R)$. It also reproduces 
accurately, but not exactly \cite{PW, Marek}, the form of 
the Keplerian angular velocity and the form of the radial 
epicyclic frequency.

In this article we address the possibility of introducing a 
pseudo-Newtonian potential in theories of gravity 
alternative to Einstein gravity. For simplicity, we limit 
ourselves to spherically symmetric spacetimes (spherically 
symmetric pseudo-potentials are often used even for 
rotating black holes). Since the field equations of the 
theories of gravity are used only to provide black hole 
solutions, it is straightforward 
to generalize to arbitrary spherically symmetric 
static metrics
the pseudo-Newtonian potential introduced by Paczynski and 
Wiita \cite{PW} for the Schwarzschild spacetime, following 
the pedagogical derivation of Ref.~\cite{Marek}. A recent 
derivation of a pseudo-Newtonian potential for certain  
spherically symmetric black hole metrics in 
Ref.~\cite{TejedaRosswog14} produced a different 
pseudo-Newtonian potential. Although the difference becomes 
irrelevant asymptotically far away from the inner edge of 
the accretion disk, it persists close to it, showing that 
the association of a pseudo-Newtonian potential with a 
black hole spacetime is not unique.

We use units in which Newton's constant $G$ and the speed 
of light in vacuo $c$ are unity, and we follow the notation 
and conventions of Ref.~\cite{Wald}. The symbol $R$ denotes 
the areal radius of spherically symmetric geometries.

\section{The Paczynski-Wiita potential for any static 
spherical black hole}
\label{sec:2}

Here we derive the analogue of the Paczynski-Wiita 
pseudo-Newtonian potential \cite{PW} for {\em any} static 
and spherically symmetric metric. We follow step-by-step 
the pedagogical derivation of Ref.~\cite{Marek}. In a 
restricted class of static spherically symmetric 
spacetimes, a different pseudo-Newtonian potential was 
obtained in Ref.~\cite{TejedaRosswog14} by studying the 
same equations for timelike geodesics (see below). 

Any static and spherically symmetric metric can be written  
in the form 
\begin{equation} \label{metric1}
ds^2 = g_{00} (R) dt^2 + g_{11} (R) dR^2 
+ R^2 d\Omega^2_{(2)} 
\end{equation}
in polar coordinates $\left( t, R, \theta, \varphi 
\right)$, where $R$ is the areal radius and 
$d\Omega_{(2)}^2=d\theta^2 +\sin^2 \theta \, d\varphi^2$ 
is 
the metric on the unit 2-sphere. The timelike and spacelike 
Killing vectors  
 $ \xi^a = \left( \partial/ \partial t 
\right)^a $ and $ \psi^a = 
\left(  \partial/ \partial \varphi \right)^a $ 
are associated with the time and rotational 
symmetries, respectively. Let  $u^a$ be the 4-tangent to a 
timelike geodesic followed by a particle of mass $m$, then  
$\xi_a u^a =u_0=-E $ and $\psi_a 
u^a= u_3 = R^2 u^3$ are  
constants of motion along these geodesics, corresponding 
to conservation of energy and angular 
momentum (per unit mass).  
Because of spherical symmetry, the orbits of test 
particles are planar and, without loss of generality, 
we assume that they take place in the $\theta = 
\pi / 2$ plane, so that $u^2 = 0$. The 
normalization of the 4-velocity gives
\begin{equation}
g_{00} (u^0)^2 + g_{11} (u^1)^2  + g_{33} (u^3)^2 = -1 \,,
\ee
or
\begin{equation}
g^{00} (u_0)^2 + g^{11} (u_1)^2 + g^{33} (u_3^2) = -1 \,.
\ee
Now, setting \cite{Marek}
\begin{equation} \label{Marek}
 g_{11} (u^1)^2 \equiv v^2 \ll 1 \,,
\ee
where $v$ is the radial velocity, we have
\begin{eqnarray}
1+v^2 &=& -g^{00} (u_0)^2 - g^{33} (u_3)^2 \nonumber\\
&&\nonumber\\
&=& (u_0)^2 \left( - g^{00} - g^{33} L^2 \right) 
\end{eqnarray}
with $L \equiv u_3/u_0 =$~const. Taking the 
logarithm gives 
\begin{equation} 
\ln \left(1 +v^2 \right)= 2\ln u_0 + \ln \left( 
-g^{00}-g^{33}L^2 \right)
\ee
and, expanding for small $v$, 
\begin{equation}\label{conservation}
\ln u_0= \frac{v^2}{2} - \frac{1}{2} \ln 
\left( - g^{00} - g^{33} L^2 \right) \,,
\ee
where $\ln u_0 = $~const. Eq.~(\ref{conservation}) has the 
form of an energy conservation equation  $ 
v^2/2  + U(R)=E/m $, where
\begin{equation}
U(R) = - \frac{1}{2} \ln \left( - g^{00} - \frac{L^2}{R^2} 
\right) \,.
\ee
Stable and unstable  circular orbits are located at the 
extrema of this potential where 
\begin{equation}
\frac{dU}{dR} = \frac{1}{g^{00} + \frac{L^2}{R^2}} 
\left\{ - \frac{1}{2} \frac{d}{dR} \left( g^{00} \right) 
+ \frac{L^2}{R^3} \right\} =0\,.
\ee
The last equation is equivalent to 
\begin{equation}
\frac{d\Phi}{dR} - \frac{L^2}{R^3} =0 
\ee
where $\Phi (R)$ is the sought-for pseudo-Newtonian 
potential, which is defined up to an irrelevant additive 
constant. The choice 
\begin{equation}
\label{generalPhi}
\Phi(R) = \frac{1}{2} \left( 1+ g^{00} \right) = 
\frac{1}{2} \left(1+ \frac{1}{g_{00} (R)} \right) 
\ee
reproduces the Paczynski-Wiita pseudo-Newtonian potential 
for the  Schwarzschild metric, which has  
$-g_{00} =g_{11}^{-1}= 1- 2m/R$ \cite{PW}:
\begin{equation} \label{PW}
\Phi_\text{{\small PW}}= \frac{- m}{R - 2m} \,.
\ee 
Eq.~(\ref{generalPhi}) gives trivially
\begin{equation}\label{trivial}
g_{00}=-\, \frac{1}{1-2\Phi}
\ee
which, in the weak-field limit $| \Phi | \ll 1$ would yield 
$g_{00}=-\left( 1+2\Phi \right)$, a relation familiar from 
the 
post-Newtonian limit of General Relativity ({\em e.g.}, 
\cite{Wald}), 
but this limit is not appropriate here because the goal is 
to investigate strong gravity near black hole horizons. 
This aspect brings us to the dichotomy inherent in the use 
of pseudo-Newtonian potentials: one wants to explore strong 
gravity, but doing this in a Newtonian way, and 
Newton's theory is intrinsically linear and 
limited to the weak-field regime.  This procedure  
apparently entails a contradiction.\footnote{It is also 
obvious that this pseudo-Newtonian potential cannot be a truly 
Newtonian description of gravity because, in vacuo ({\em 
e.g.}, for the Schwarzschild geometry), it should then be 
$\nabla^2 \Phi=0$, while in general, this Laplacian does 
not vanish. Replacing the flat space Laplacian with 
the Laplace-Beltrami operator $g^{ij}\nabla_i \nabla_j \Phi 
$ ($i,j =1,2,3$) does not 
help, either.} However, pseudo-Newtonian 
potentials do not attempt to describe {\em all} aspects of 
physics in the strong gravity regime, but only to catch 
certain aspects, {\em i.e.}, the innermost stable and 
outermost marginally stable circular orbits. 
Tejeda and Rosswog \cite{TejedaRosswog14} define a 
pseudo-Newtonian potential for  static, 
spherically symmetric metrics of the form~(\ref{metric1}) 
which, in addition, satisfy the condition 
$g_{00} \, g_{11}=-1$. Their pseudo-potential 
is 
\begin{equation} 
\label{TejedaRosswog}
\Phi_\text{{\small TR}}(R)= -\frac{\left( g_{00}+1 
\right)}{2}
\ee
which, using eq.~(\ref{trivial}), relates to our 
potential~(\ref{generalPhi}) through
\begin{equation}\label{comparison}
\Phi_\text{{\small TR}}=\frac{\Phi}{1-2\Phi} \,.
\ee
It is only in the limit $|\Phi| \ll 1$ that the two 
pseudo-potentials coincide and we have already discussed 
how this limit is not appropriate in the vicinity of a 
black hole. That the pseudo-Newtonian potentials $\Phi$ and 
$\Phi_\text{{\small TR}}$ do not coincide is best seen at 
the horizon: when $g_{00} \rightarrow 0^{-}$, $\Phi 
\rightarrow 
-\infty$ while $\Phi_\text{{\small TR}} \rightarrow -1/2$. 
Apart from a sign, the difference resides basically in  
the fact 
that the inverse metric coefficient  $g^{00}$ appears in 
the generalized Paczynski-Wiita 
potential~(\ref{generalPhi}), while $g_{00}$ appears in the 
Tejeda-Rosswog potential~(\ref{TejedaRosswog}). Sarkar, 
Ghosh, and Bhadra \cite{SarkarGhoshBhadra2014} 
introduce a velocity-dependent potential to include 
special-relativistic effects in the dynamics, by 
considering spherically symmetric static metrics with the 
usual restriction $g_{00}\, g_{11}=-1$. When the velocity 
terms are dropped, their potential reduces 
to~(\ref{TejedaRosswog}). 

The lesson to draw from eq.~(\ref{comparison}) 
is that, at least in principle if not in practice, the 
pseudo-Newtonian potential associated with a given 
spacetime metric is not unique.

At this point, it is 
appropriate to  make explicit the assumptions used to 
derive eq.~(\ref{generalPhi}):
\begin{itemize}
\item The spacetime metric $g_{ab}$ is static and 
spherically symmetric.
\item The metric is written in the gauge~(\ref{metric1}) 
using the areal radius $R$ (which is defined in a 
geometric, coordinate-independent way).
\item The radial velocities $v$ of the massive test 
particles  defined by $v^2 \equiv g_{11} (u^1)^2 $ are 
small everywhere along the timelike geodesics in 
comparison with the speed of light (the tangential velocities, 
by contrast, are not restricted to be small).
\item The pseudo-Newtonian potential is required to 
produce at its extrema the circular orbits of 
the spacetime geometry~(\ref{metric1}).
\end{itemize}

In particular, the Einstein equations have not been used 
and the formula~(\ref{generalPhi}) is valid in any theory 
of gravity in which massive test particles follow timelike 
geodesics. What is more, asymptotic flatness of the 
metric~(\ref{metric1}) is not required (see below). 

The fact that the result~(\ref{generalPhi}) does not 
depend on the theory of gravity (provided that 
test particles follow geodesics)  will be useful to study 
particle trajectories and accretion around black holes in 
alternative theories of gravity. Before approaching this 
problem, however, it is useful to restrict to  General 
Relativity and give a geometric characterization of the 
pseudo-Newtonian potential~(\ref{generalPhi}) obtained and 
some examples.

\section{General Relativity}
\label{sec:3}

In this section we restrict ourselves to General 
Relativity and we assume that the geometry is described by 
eq.~(\ref{metric1}). 

\subsection{Relation with the Misner-Sharp-Hernandez mass 
in General Relativity}

In this subsection we make 
the further assumption that 
\begin{equation}\label{Jacobson}
g_{00} \, g_{11}=-1 \,.
\ee
This assumption encompasses a wide class of spherically 
symmetric geometries:\footnote{Early interest in these 
metrics 
includes Refs.~\cite{BondiKilmister60, French77}.}  the 
condition~(\ref{Jacobson}) has been 
studied in Ref.~\cite{Jacobson}, where it is shown that it 
is equivalent to require that the (double) projection of 
the 
Ricci tensor onto radial null vectors $l^a$ vanishes, 
$R_{ab}l^a l^b =0$. This condition is also equivalent to 
require 
that the restriction of the Ricci tensor to the $\left( 
t,R\right)$ subspace is proportional to the restriction of 
the metric $g_{ab}$ to this subspace \cite{Jacobson}. 
Equivalently, the areal radius $R$ is an affine parameter 
along radial null geodesics \cite{Jacobson}. These results 
hold in higher spacetime dimension as well, and 
the geometries satisfying the condition~(\ref{Jacobson}) in 
General Relativity include vacuum, electrovacuum with 
either Maxwell or 
non-linear Born-Infeld electrodynamics, and a spherical 
global monopole (``string hedgehog'' \cite{hedgehog}) 
\cite{Jacobson}.
Under this assumption, we can 
use the characterization of the Misner-Sharp-Hernandez mass 
$M_\text{{\small MSH}}$ \cite{MSH} 
\begin{equation} 
1-\frac{2M_\text{{\small MSH}}}{R}=\nabla^cR \nabla_c R 
\ee and the 
fact that, in the coordinates $\left( t,R, \theta, \varphi 
\right)$ used, $\nabla^c R \nabla_c R=g^{RR}$ to obtain 
\begin{equation} \label{characterization}
\Phi(R)= -\, \frac{M_\text{{\small MSH}}(R) }{R 
-2M_\text{{\small MSH}}(R) } \,,
\ee 
where $M_\text{{\small MSH}}(R)$ is the 
Misner-Sharp-Hernandez mass 
contained in a 2-sphere of symmetry of radius $R$. This is 
a quasilocal 
mass: the Hawking-Hayward quasilocal energy 
\cite{HawkingHayward}, which seems to be favoured by 
the relativity community among the 
various notions of quasilocal energy introduced in General 
Relativity since the 1960s (see the review 
\cite{Szabados}), is well known to reduce to the 
Misner-Sharp-Hernandez mass in spherical symmetry 
\cite{Haywardspherical}. Since both 
the areal radius and the Misner-Sharp-Hernandez mass are 
geometric quantities defined in a coordinate-independent 
way, this formula constitutes a geometric characterization 
of the generalized Paczynski-Wiita 
potential~(\ref{generalPhi}). Moreover, 
eq.~(\ref{characterization}) generalizes eq.~(\ref{PW}) 
valid for the Schwarzshild geometry. The use of the 
Misner-Sharp-Hernandez mass to express the pseudo-Newtonian 
potential is particularly appropriate because this 
construct has a Newtonian character, contrary to other 
quasilocal energies in the literature \cite{BlauRollier, 
myNewtHH}.

The (apparent) horizons in spherical symmetry are located at 
the roots of the equation $\nabla_cR \nabla^c R=0$ 
(see, {\em e.g.}, \cite{mylastbook}) and therefore the 
pseudo-potential $\Phi$ diverges at the (apparent) black 
hole 
horizon of radius $R_\text{{\small AH}}$, where 
$R_\text{{\small AH}}=2M_\text{{\small 
MSH}}(R_\text{{\small AH}})$ which is, 
of 
course, reminiscent of the behaviour of the Paczynski-Wiita 
potential at the Schwarzschild horizon. If, in addition, 
asymptotic flatness is imposed, the metric coefficient 
$g_{00} \rightarrow -1$ as $R\rightarrow +\infty$ and $\Phi 
\rightarrow $~const. in this limit. However, it is not 
necessary to impose asymptotic flatness and in certain 
situations it may even be inappropriate ({\em e.g.}, in the 
Schwarzschild-de Sitter-Kottler solution). For example,  
alternative theories of gravity designed to explain the 
present acceleration of the universe without invoking an 
{\em ad hoc} dark energy contain an effective time-dependent 
cosmological ``constant'' and black hole solutions are not 
asymptotically flat in these theories, but asymptotically 
Friedmann-Lema\^itre-Robertson-Walker \cite{reviews}. 

The spherically symmetric metric considered in 
Ref.~\cite{TejedaRosswog14} satisfies the condition $g_{00} 
\, g_{11}=-1$. It is, therefore, possible to express the 
Tejeda-Rosswog potential~(\ref{TejedaRosswog}) using the 
Misner-Sharp-Hernandez mass. The result is 
\begin{equation}
\Phi_\text{{\small TR}}= -\frac{M_\text{{\small MSH}} }{R} 
\,,
\ee
which shows again the crucial difference between $\Phi$ and 
$\Phi_\text{{\small TR}}$ at the horizon 
$R=2M_\text{MSH}(R)$.

The Misner-Sharp-Hernandez mass \cite{MSH} (or, more in 
general, the Hawking-Hayward mass \cite{HawkingHayward}) is 
only defined in General Relativity and in Lovelock gravity 
\cite{HidekiLovelock}, therefore   
eq.~(\ref{characterization}) does not apply to other 
theories of gravity, while eq.~(\ref{generalPhi}) does.

\subsection{Examples in Einstein theory}

We have already discussed how eq.~(\ref{generalPhi}) 
reproduces the original Paczynski-Wiita potential for the  
Schwarzschild black hole. Let us consider now other 
examples in Einstein theory.

\subsubsection{Schwarzschild-de Sitter-Kottler black hole}
A second example is given by the 
Schwarzschild-de Sitter-Kottler geometry, which can be 
written in static coordinates as
\begin{eqnarray}
ds^2 &=& -\left( 1-\frac{2m}{R}-H^2R^2 \right) dt^2 
+ \frac{dR^2}{ 1-\frac{2m}{R}- H^2R^2} \nonumber\\
&&\nonumber\\
&\, & +R^2  d\Omega_{(2)}^2  \,, \label{SdS}
\end{eqnarray}
where $H_0=\sqrt{\Lambda/3}$ is the constant Hubble 
parameter and $\Lambda>0$ is the cosmological constant. The 
locally static metric is already in the 
form~(\ref{metric1}) with 
$g_{00} \, g_{11}=-1$ in the region between the 
cosmological 
and black hole horizons, and the 
corresponding pseudo-Newtonian 
potential is
\begin{equation}\label{PhiSdS}
\Phi_\text{{\small SdS}}(R)= - \, \frac{ 
\left( \frac{m}{R}+\frac{H^2R^2}{2}\right) }{
1-\frac{2m}{R}-H^2R^2} 
\,,
\ee
which coincides with the pseudo-Newtonian potential for 
Schwarzschild-de Sitter found in Refs.~\cite{SdS1, SdS2}. 
This example shows how asymptotic flatness is not a 
requirement for the pseudo-Newtonian 
description~(\ref{generalPhi}). This fact has some 
importance because, as noted above, in theories of 
modified gravity designed to explain the present 
acceleration of the universe without dark energy, black 
holes are asymptotically 
Friedmann-Lema\^itre-Robertson-Walker.  Currently, there is 
much theoretical effort devoted to predicting the effects 
of scalar (and other) hair around black holes in modified 
gravity \cite{scalarhair}, which makes them non-isolated.

\subsubsection{Kiselev black hole}

The Kiselev solution of the Einstein equations 
\cite{Kiselev} describes  
a static spherical black hole surrounded by 
$n$ non-interacting quintessence fluids. It can be used 
as a toy model to study the qualitative modifications 
induced on a black hole by a dark energy environment.
The line element is \cite{Kiselev}
\begin{eqnarray}
ds^2 &=&- \left[ 1 - \frac{2m}{R} - \sum_{j=1}^n \left( 
\frac{R_j}{R}\right)^{3w_j +1} \right] dt^2 \nonumber\\
&&\nonumber\\
&\, &  +  \frac{dR^2}{ 
1 - \frac{2m}{R} - \sum_{j=1}^n \left( \frac{R_j}{R} 
\right)^{3w_j +1} } +R^2 d\Omega_{(2)}^2 \,,
\end{eqnarray}
where the equation of state parameters of the $n$ fluids 
satisfy  $-1<w_j<-1/3$ ($j=1, \, ... \, n$). This solution 
has a black hole horizon at the roots of the equation 
$g^{RR}=0$ (when these exist).

Using eq.~(\ref{generalPhi}), the pseudo-Newtonian 
potential for the Kiselev black hole is found to be
\begin{equation}
\Phi_\text{{\small Kiselev}} =  - \frac{2m + \sum_{j=1}^n 
\left( 
\frac{R_j^{3w_j+1}}{R^{3w_j}} \right) }{2 \left[ R - 2m - 
\sum_{j=1}^n \left( \frac{R_j^{3w_j+1}}{R^{3w_j}} \right)  
\right] } \,,
\ee
or, for a single fluid,
\begin{equation}\label{Ksinglefluid}
\Phi_\text{{\small Kiselev}} = - \frac{\left[ m + 
\frac{R}{2} \left( 
\frac{R_0}{R} \right)^{3w+1}  \right]}{ R - 2m - R 
\left( \frac{R_0}{R} \right)^{3w+1}} \,.
\ee
Ref.~\cite{Kovar} derives the pseudo-Newtonian potential 
for the special case of a single fluid with $w=-2/3$ 
(although the generalization to any value of $w$ in the 
interval $\left( -1, -1/3 \right)$ is straightforward), and 
the 
result coincides with eq.~(\ref{Ksinglefluid}) specialized 
to this single fluid. The 
Schwarzschild-de Sitter-Kottler solution~(\ref{SdS}) 
and the corresponding pseudo-Newtonian 
potential~(\ref{PhiSdS}) are recovered in the parameter 
limit $w \rightarrow -1$ for a single fluid. 

The determination of marginally stable orbits (innermost 
stable circular orbit and outermost circular orbit) for the 
Kiselev solution is not trivial because it involves solving 
a quintic equation for the radii of these orbits 
\cite{Kovar, Onoetal15}. The Sturm theorem was applied to 
this problem, determining the condition for the existence 
of these orbits, in Ref.~\cite{Onoetal15}.

\section{Pseudo-Newtonian potential in modified gravities}
\label{sec:4}

Currently, there is significant theoretical research 
devoted to probing and testing gravity using black holes in 
view of the Event Horizon Telescope aiming at resolving the 
surroundings of black holes up to 1.5 Schwarzschild radii 
({\em e.g.}, \cite{EHT}). Therefore, it is appropriate to 
explore solutions of alternative theories of gravity to 
obtain some insight on the qualitative deviations of black 
hole properties from those of General Relativity. Physical 
observables are the size of the accretion disks around 
black holes, which are characterized by the innermost and 
the outermost stable orbits, and the frequency of the 
radiation emitted near the inner edge of the accretion 
disk.

\subsection{Modified Schwarzschild black holes in 
quadratic gravity}

A formalism incorporating  small deviations from a 
Schwarzschild black hole in a wide class of quadratic 
theories of gravity was proposed in 
Ref.~\cite{YunesStein2011}. 
The action includes dynamical Chern-Simons gravity as a 
special case and is
\begin{eqnarray}
S &=& \int d^4x \sqrt{-g} \left[ \frac{{\cal R}}{16\pi G} 
+\alpha_1 f_1(\phi) {\cal R}^2 
+\alpha_2 f_2(\phi) R_{ab}R^{ab} \right. \nonumber\\
&&\nonumber\\
&\, & \left. +\alpha_3 f_3(\phi) R_{abcd} R^{abcd} 
+\alpha_4 f_4(\phi)  R_{abcd} {}^*R^{abcd} 
\right.\nonumber\\
&&\nonumber\\
&\, & \left. -\frac{\beta}{2} \nabla^c\phi 
\nabla_c\phi   +{\cal L}^\text{(m)} \right] \label{action}
\end{eqnarray}
where ${\cal R}$ is the Ricci curvature of the metric 
$g_{ab}$ with determinant $g$, $R_{ab}, R_{abcd}$, and 
${}^*R_{abcd}$ are the Ricci and Riemann tensor and the 
dual of the latter, respectively, $\phi$ is a scalar field, 
$\alpha_i$ and $ \beta$ are coupling constants and 
$f_i(\phi)$ ($i=1, \, ... \, ,4$) are coupling functions 
which can be 
Taylor-expanded around $\phi=0$. The spherically 
symmetric, static, and 
asymptotically flat  geometry 
describing deviations from the Schwarzschild metric is found 
to be \cite{YunesStein2011}
\begin{eqnarray}
ds^2 &=& - \left( 1-\frac{2M}{R}\right) \left( 1+h(R) 
\right) dt^2 
+\frac{1+k(R)}{1-\frac{2M}{R}} \, dR^2 \nonumber\\
&&\nonumber\\
& \, & +R^2 d\Omega_{(2)}^2 \,, \label{YunesStein}
\end{eqnarray}
where
\begin{eqnarray}
h(R) &=& \frac{ \zeta}{3(1-2M/R) } \, \left( \frac{M}{R} 
\right)^3 \tilde{h}(R) \,, \\
&&\nonumber\\
k(R) &=& \frac{ -\zeta}{1-2M/R} \, \left( \frac{M}{R} 
\right)^2 \tilde{k}(R) \,,\\
&&\nonumber\\
\zeta &=& \frac{16\pi \alpha_3^2}{\beta M_0^4} \,,
\end{eqnarray}
and where $M_0$ is a bare mass related to the physical mass 
$M$ by
\begin{equation}
M=M_0 \left( 1+ \frac{49\zeta}{80} \right) \,,
\ee
which leads to the implicit equation for the 
dimensionless function $\zeta(M)$ \cite{YunesStein2011}  
\begin{equation}
\zeta \left( 1+\frac{49\zeta}{80} \right)^4 =\frac{16\pi 
\alpha_3^2}{\beta M^4} \,.
\ee
The functions $\bar{h}(R)$ and $ \bar{k}(R)$ are given by 
\cite{YunesStein2011}
\begin{eqnarray}
\tilde{h}(R) &=& 1+ \frac{26M}{R} 
+\frac{66M^2}{5R^2}
+\frac{96M^3}{5R^3}
-\frac{80M^4}{R^4} \,,\\
&&\nonumber\\
\tilde{k}(R) &=& 1+ \frac{M}{R} 
+\frac{52M^2}{3R^2}
+\frac{2M^3}{R^3}
+\frac{16M^4}{5R^4} 
-\frac{368M^5}{3R^5} \,.\nonumber\\
&& 
\end{eqnarray}
The solution~(\ref{YunesStein}) is 
regular everywhere except at $R=0$ 
where it exhibits the usual black hole spacetime 
singularity. Eq.~(\ref{generalPhi}) 
then yields the pseudo-Newtonian 
potential in the wide class of quadratic theories
of gravity~(\ref{action})
\begin{equation}
\Phi(R)= \frac{-M}{R} \, 
\frac{1- \frac{\zeta}{6\left( 1-2M/R\right)} \left( 
\frac{M}{R} \right)^2 \left(1-2M/R\right) \tilde{h}(R) }{
\left( 1-2M/R \right) \left[ 1+ \frac{3}{\left( 1-2M/R 
\right)} \left( \frac{M}{R} \right)^3 \tilde{h}(R) \right]} 
\,.
\ee
The radius of the innermost stable circular orbit 
corresponding to 
$\Phi'(R)=0$ is computed in \cite{YunesStein2011} as
\begin{equation}
R_\text{{ \small ISCO}}=\left( 
6-\frac{16297\zeta}{9720}\right) M \,.
\ee
In this case the spacetime  metric does not satisfy the 
condition $g_{00} \, g_{11} =-1$  and, in the weak-field 
limit, 
there are two post-Newtonian potentials since the 
post-Newtonian line element has the form
\begin{equation}
ds^2=-\left( 1-2\Phi \right) dt^2 + \left( 1+2\Psi \right) 
dR^2 +R^2 d\Omega_{(2)}^2 \,.
\ee
This feature is well known in alternative theories of 
gravity, also for cosmological perturbations. The fact that 
the pseudo-Newtonian potential apparently depends only on 
$g_{00}$  (and not on $g_{11}$) seems to be a limitation of 
attempts to probe alternative theories of gravity near 
black hole horizons which use such pseudo-potentials. 
However, this is not entirely true since $v^2$  depends 
also on $g_{11}$, as shown by eq.~(\ref{Marek}). This 
aspect is discussed in the next subsection.

\subsection{Epicyclic frequency}

Consider now orbits in the equatorial plane $z=0$. 
Switching from spherical to cylindrical coordinates $\left( 
r, \varphi, z \right)$, the 
effective potential and the 
pseudopotential~(\ref{generalPhi}) are  cylindrically 
symmetric, $U= U(r, z)$ and $\Phi=\Phi(r,z)$. 
Consider an equatorial circular orbit at an extremum of the 
effective potential $U$, given by $\left( r_0, \varphi_0 
+\Omega t, 0 \right)$ (where $\varphi_0$ is an azimuthal 
initial condition) and  constant angular momentum 
$L_{(0)}$. The Keplerian frequency 
$\Omega=\dot{\varphi} $ of this orbit satisfies ({\em 
e.g.}, \cite{BinneyTremaine}) 
\be
\Omega^2 = \left. \frac{1}{r} \, \frac{\partial 
\Phi}{\partial r} \right|_{\left( r_0, 0 \right)}  
=\frac{L_{(0)}}{r_0^2} 
\,.
\ee
In the case of the potential~(\ref{generalPhi}) we have
\be
\Omega^2= \left. \frac{1}{2r} \, \frac{ \partial 
g^{00}}{\partial r} 
\right|_{\left( r_0, 0 \right)}  \,.
\ee
Now perturb this circular orbit so that
\begin{eqnarray}
r(t) &=& r_0 +\delta r(t) \,,\\
&&\nonumber\\ 
\varphi(t) &=&\varphi_0 +\Omega t + \delta \varphi(t) \,,\\
&&\nonumber\\
z(t) &=& \delta z(t) \,,
\end{eqnarray}
and the angular momentum per unit mass is $L=L_{(0)}+\delta 
L$. In the approximation of small perturbations, the horizontal
and vertical  epicyclic frequencies  $\kappa$ and 
$\nu$ are  given by 
\begin{eqnarray}
\kappa^2 &=& \left[ \frac{\partial^2 \Phi}{\partial r^2} 
+\frac{3L_{(0)}^2}{r^4} \right]_{\left( r_0, 0 \right)} \\
&&\nonumber\\
&=& 
\left.\frac{2\Omega}{r} \, \frac{\partial }{\partial r} 
\left( 
\Omega^2 r \right) \right|_{\left( r_0, 0 \right)} = \left. 
4\Omega^2 +r\, \frac{\partial (\Omega^2)}{\partial r} 
\right|_{\left( r_0, 0 \right)} \,,
\nonumber\\
&&\nonumber\\
\nu^2 &=& \left. \frac{\partial ^2 \Phi}{\partial z^2} 
\right|_{\left( r_0, 0 \right)} \,,
\end{eqnarray}
respectively \cite{BinneyTremaine}. In our case these 
formulae give
\begin{eqnarray}
\kappa^2 &=& \left. \frac{1}{2} \, \frac{\partial^2 
g^{00}}{\partial r^2} 
 \right|_{\left( r_0, 0 \right)} \,,\\
&&\nonumber\\
\nu^2 &=& \left. \frac{1}{2} \, \frac{\partial^2 
g^{00}}{\partial z^2} \right|_{\left( r_0, 0 \right)} \,,
\end{eqnarray}
In the case of a stable circular orbit of radius $r_0$, a 
test particle will oscillate with $\delta r 
=\delta x^1= \delta_0 \cos \left( \kappa t \right)$. 

As for the radius of the innermost stable circular orbit, 
it seems that the horizontal and vertical epicyclic 
frequencies depend 
only on $g_{00}$ and not on other metric components.
However, the quantity $v$ appearing in the  
conservation equation $ v^2/2  + U(R)=E/m $ of the previous 
section is related with $g_{RR}$ by eq.~(\ref{Marek}). This 
relation can be interpreted as saying  that, if  
{\em proper} (instead of coordinate) radii and vertical 
distances are to be used, then one must replace the  
intervals $dr$ and $dz$ with  
$dr_{\mbox{prop}}=\sqrt{g_{rr}} \, dr$ and 
$dz_{\mbox{prop}}=\sqrt{g_{zz}} \, dz$.  This replacement 
is consistent with the definition of the radial velocity 
$v^2 \equiv g_{11} \left( u^1 \right)^2 $ in \cite{Marek}. 
Then, the expressions of the Keplerian and epicyclic 
frequencies $\Omega$, $\kappa$ and $\nu$ should be replaced 
by $\Omega/\sqrt{g_{rr}}$, $\kappa/\sqrt{g_{rr}}$, and 
$\nu/\sqrt{g_{zz}}$, making these physical quantities 
dependent on metric components other than $g_{00}$. 
Then, testing particle orbits would mean doing  more 
than merely testing gravitational shifts. Remember, however, 
that the Paczynski-Wiita potential does not reproduce the epicyclic 
frequencies accurately even in the Schwarzschild spacetime \cite{Marek}.

Let us discuss now the condition~(\ref{Marek}) in 
cylindrical coordinates in the equatorial plane.  Consider 
the coordinate transformation 
\be
\left\{ x^{\mu} \right\}= \left( t,  r, \varphi, z \right) 
\rightarrow \left\{ x^{\mu '}\right\} =  
\left( t,R, \theta, \varphi \right) 
\ee
with
\begin{eqnarray}
R &=& \sqrt{r^2 +z^2} \,, \label{R(rz)}\\
&&\nonumber\\
\varphi &=& \varphi \,,\\
&&\nonumber\\
\theta &=& \tan^{-1} \left( \frac{r}{z} \right) \,,
\end{eqnarray}{
and inverse
\begin{eqnarray}
r &=& R \sin\theta  \,,\\
&&\nonumber\\
\varphi &=& \varphi \,,\\
&&\nonumber\\
z &=& R\cos \theta  \,.
\end{eqnarray}
Eq.~(\ref{Marek}) states that $v^2 = g_{RR} (u^R)^2$, where 
$u^R \equiv dR/d\tau$ and  $\tau$ is the proper time  
along a timelike geodesic. Using the transformation 
property of the 
metric tensor 
\be
g_{\mu ' \nu '}=\frac{ \partial x^{\mu}}{\partial x^{\mu 
'}} \, \frac{ \partial x^{\nu}}{\partial x^{\nu'}} \, 
g_{\mu\nu} 
\ee
one obtains
\begin{eqnarray}
g_{RR} &=& \frac{ \partial x^{\mu}}{\partial R} \, 
 \frac{ \partial x^{\nu}}{\partial R} \,g_{\mu\nu} 
\nonumber\\
&=& g_{rr}\left( \frac{ \partial r}{\partial R} \right)^2 
+g_{rr}\left( \frac{ \partial z}{\partial R} \right)^2 
+2 g_{rz} \frac{ \partial z}{\partial R} \, \frac{ \partial 
r}{\partial R}
\nonumber\\
&&\nonumber\\
&=& \sin^2 \theta g_{rr} +\cos^2 \theta g_{zz}+ \sin 
(2\theta) g_{rz} \,.
\end{eqnarray}
On the equatorial plane $\theta=\pi/2$, it is
$ g_{RR}=g_{rr} $.  We now have 
\be
u^R \equiv \frac{dR}{d\tau}= \frac{1}{R} \left( ru^r +z u^z 
\right)
\ee
using eq.~(\ref{R(rz)}) and, on the equatorial plane, 
$u^R=u^r$.  Therefore, it is also 
\be
 v^2=g_{RR} (u^R)^2= g_{rr} (u^r)^2
\ee 
on the $z=0$ plane. 
The radial velocity near an equatorial circular orbit 
then obeys 
\be 
\delta \dot{r} \equiv \frac{d \delta r}{dt}= 
\frac{dr}{d\tau}\, \frac{d\tau}{dt}
=\frac{\delta u^r}{u^0}= \frac{ v }{\sqrt{g_{rr}}  u^0} 
\simeq \frac{v}{\sqrt{g_{rr}} } 
\ee
to first order. Since $v$ depends on $g_{rr}$, which 
is in general distinct from $g_{00}$ in alternative 
theories of gravity, there is hope for tests of 
black  hole metrics which solve these theories.

\section{Conclusions}
\label{sec:5}

Given the many reasons to study theories of gravity 
alternative to General Relativity and, as a consequence, 
black holes in these theories, and given that the future 
Event Horizon Telescope promises to image the black hole 
SgrA$^*$ at the centre of our galaxy with a resolution of 
1.5 Schwarzschild radii \cite{EHT}, it is of interest to 
study the dynamics of particles around black holes in 
general spacetimes, not only in those (Schwarzschild or 
Kerr) which solve the Einstein equations of General 
Relativity. A simplifying tool widely used in numerical 
simulations of accretion disks around black holes is the 
pseudo-Newtonian potential. Restricting, for simplicity, to 
static and spherically symmetric black hole metrics, we 
have shown how a pseudo-Newtonian potential can be derived 
for any such spacetime, following step-by-step the 
pedagogical derivation of the Paczynski-Wiita potential 
\cite{PW} given in Ref.~\cite{Marek}. The generalization of 
the Paczynski-Wiita potential to any static spherically 
symmetric metric (not necessarily representing a black 
hole\footnote{Ref.~\cite{SarkarGhoshBhadra2014} indeed 
derives a pseudo-Newtonian potential for naked 
singularities.}) is straightforward, but nevertheless it 
comes with a surprise. A previous generalization of this 
potential, restricted to the subclass of static spherical 
symmetric metrics satisfying the condition $g_{00} \, 
g_{11}=-1$ \cite{TejedaRosswog14}, produces a different 
pseudo-Newtonian potential, although essentially the same 
equations ({\em i.e.}, the equations for timelike 
geodesics) were considered in \cite{TejedaRosswog14}. There 
is no obvious compelling reason to prefer one of the two 
potentials over the other. Therefore, this discrepancy 
means that even the pseudo-Newtonian potential associated 
with a simple black hole spacetime is not unique. There are 
now many pseudo-Newtonian potential functions in the 
literature, for both spherical and cylindrically symmetric 
(rotating) black holes. Since these potentials are just 
effective quantities, mere tricks used to simplify 
complicated equations, one should not regard them as 
fundamental quantities and attribute to them more 
importance than they deserve. Nevertheless, these 
pseudo-potentials are used in practical calculations in 
astrophysics and it would be worth understanding them fully 
together with their limitations. An interesting aspect is 
the expression of the (generalized) Paczynski-Wiita 
potential in terms of the Hawking quasilocal mass (which in 
spherical symmetry reduces to the Misner-Sharp-Hernandez 
mass). It would be interesting to relate this potential, 
and other pseudo-Newtonian potentials, also to the other 
known quasilocal energy constructs if possible.

\begin{acknowledgments} 

This research is supported by Bishop's University and by 
the Natural Sciences and Engineering Research Council of 
Canada ({\em NSERC}).

\end{acknowledgments}



\end{document}